\documentclass[supplement]{ptptex}

\usepackage{graphicx}

\title{%     
Forward-backward flow correlations in relativistic heavy-ion collisions%
\footnote{Presented by WB at XLI International Symposium on
Multiparticle Dynamics (ISMD2011), 26-30 September 2011, Miyajima, Japan}}

\author{%       %Use \scshape  for the family name.
Wojciech \textsc{Broniowski},$^{1,2}$
Piotr \textsc{Bo\.zek},$^{2,3}$
Jo\~{a}o \textsc{Moreira}$^{4}$
}

\inst{%         %Affiliation, neglected when [addenda] or [errata].
$^1$Institute of Physics, Jan Kochanowski University, PL-25406~Kielce, Poland\\
$^2$The H. Niewodnicza\'nski Institute of Nuclear Physics, Polish Academy of Sciences, PL-31342 Krak\'ow, Poland\\
$^3$Institute of Physics, Rzesz\'ow University, PL-35959 Rzesz\,ow, Poland\\
$^4$Centro de F\'{i}sica Computacional, Department of Physics, University of Coimbra, 3004-516 Coimbra, Portugal
}

\abst{%   
We discuss the torque effect in the initial fireball formed in relativistic heavy-ion collisions, manifesting itself, 
on the event-by-event basis, in 
a relative angle between the principal axes of the transverse momentum distributions in the forward and backward
rapidity regions. The torque follows from two natural features: 1)~the sources of particles (e.g. wounded nucleons) 
emit predominantly in their forward hemispheres, and 2)~there exist fluctuations in the transverse 
distribution of sources from the two colliding nuclei. On the average, the 
standard event-by-event deviation of the relative torque angle is about $20^\circ$ for the central and $10^\circ$ for the mid-peripheral collisions.
The hydrodynamic expansion of a torqued fireball leads to a torqued 
collective flow of the fluid, which, in turn, yields torqued principal axes of the transverse-momentum
distributions at different rapidities.
We discuss experimental measures based on cumulants involving particles in different rapidity regions, 
which allow for a quantitative extraction of the effect from the experimental data. 
We estimate the non-flow contributions from resonance decays with the help of {\tt THERMINATOR}.
}

\begin{document}

\maketitle

The forward-backward rapidity correlations 
reveal important information on the mechanism of particle production in high-energy hadronic and nuclear collisions,
uncovering the features
of the dynamical system  at a very early stage.
This talk is based on our recent work\cite{Bozek:2010vz}, where more details as well as 
a complete list of references may be found. We discuss an interesting forward-backward effect, concerning the 
event-by-event fluctuations of the longitudinal 
shape of the fireball created in relativistic heavy-ion collisions. 

\begin{figure}[tb]
    \parbox{.45\textwidth}{%   %\def\halftext{.471\textwidth}
    \includegraphics[width=.45\textwidth]{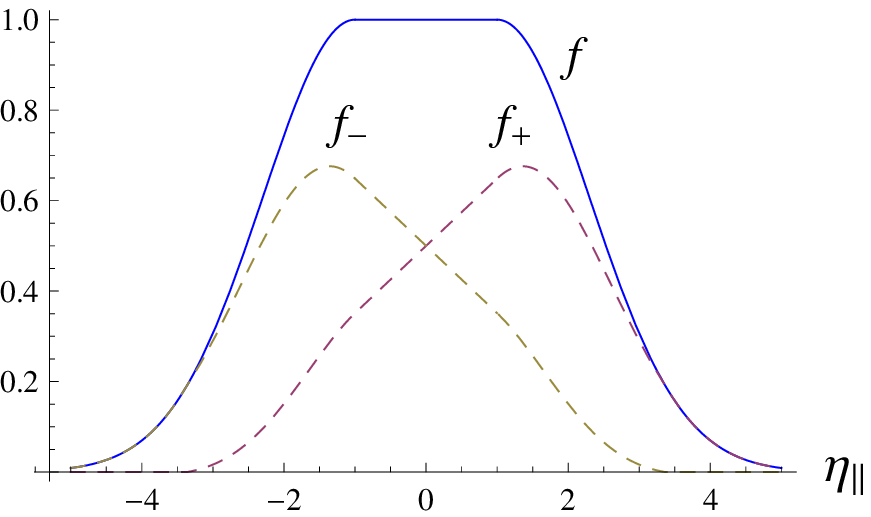}
\caption{The emission profiles in space-time rapidity $\eta_\parallel$ for the wounded nucleons 
(dashed lines) and the binary collisions (solid line). The profiles $f+$ and $f_-$ corresponds, respectively, to the 
forward and backward moving wounded nucleons.\label{fig:fplot}}}
            \hfill
     \parbox{.53\textwidth}{
     \centerline{\includegraphics[width=.28\textwidth]{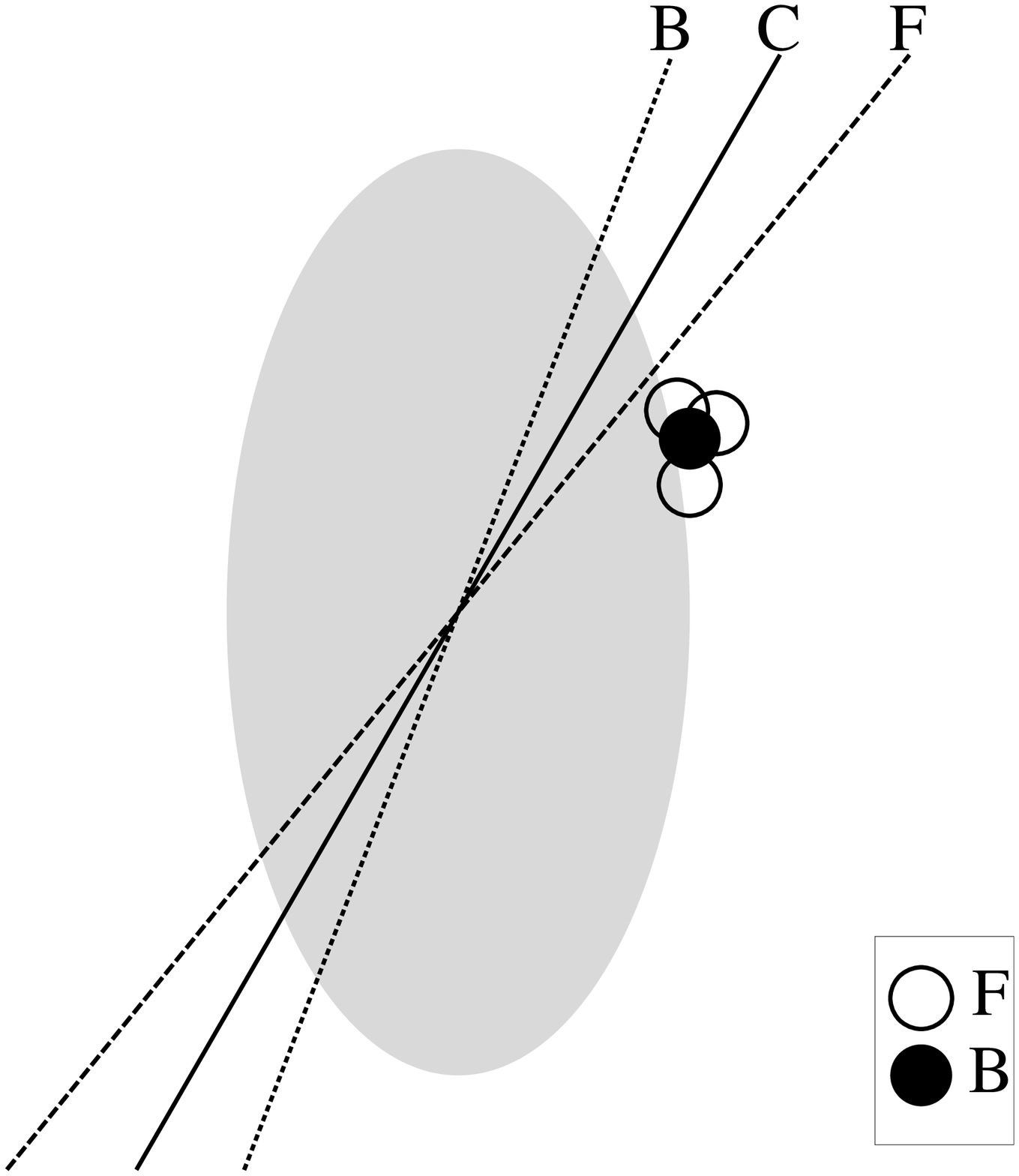}}
\caption{Emergence of the torque effect: A random cluster of wounded nucleons moving in the 
forward (F) and backward (B) directions 
causes a random torque of the principal axes. The angle is higher in the 
forward direction than in the backward direction. (C) indicates the central region.
\label{fig:simple}}}
\end{figure}

\begin{figure}[tb]
\centerline{\includegraphics[width=.4\textwidth]{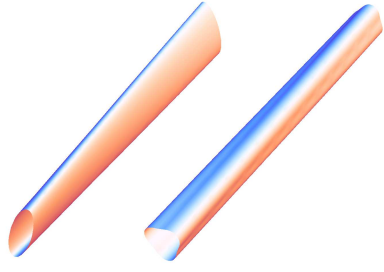}}
\caption{The schematic figure of the torqued fireball, elongated along the $\eta_\parallel$ axis. 
The direction of the principal axes in the transverse plane 
rotates as $|\eta_\parallel|$ increases. The left and right pictures correspond to the rank-2 (elliptic) and rank-3 (triangular) cases, 
respectively. The effect occurs on event-by-event basis. \label{fig:cartoon}}
\end{figure}

The effect relies on 
two basic facts: 
1)~The wounded nucleons~\cite{Bialas:1976ed} emit particles predominantly in their forward hemispheres (see Fig.~\ref{fig:fplot}) 
-- this feature is 
strongly supported with the data analyses of the \mbox{d-Au} \cite{Bialas:2004su} and AA \cite{Bzdak:2009dr} collisions, as well as 
a successful (for the first time) description of the directed flow \cite{Bozek:2010bi} at RHIC. 2)~There are random fluctuations as 
explained in Fig.~\ref{fig:simple}. The key point here is that the number of wounded nucleons in a cluster may be asymmetric, i.e., 
it may contain more nucleons from nucleus A than B. Since the emission profiles are asymmetric, the torque of the principal axes 
is higher in the direction of motion where more wounded nucleons are moving.  
As a result, the initial fireball is torqued on the event-by-event basis, as graphically shown in Fig.~\ref{fig:cartoon}.
Specifically, we use the mixed model of the initial Glauber phase, adding 14\% of binary collisions to the wounded nucleons.\cite{Bozek:2010vz}
The initial fireball is then evolved hydrodynamically (here we use perfect 3+1 dimensional hydrodynamics with a 
realistic equation of state), which leads to a torque in the collective flow velocity. This, in turn, yields 
a torque of the principal axes (between the forward and backward regions) of the measured transverse-momentum distributions of created hadrons. 
Since statistical hadronization leads to random fluctuations of the direction of principal 
axes, one needs to look at the torque effect and the prospects of its experimental observation with care. For that purpose 
we have proposed \cite{Bozek:2010vz} to consider cumulant measures involving particles from the forward and backward regions. Let
\begin{eqnarray}
\left \langle e^{i n(\phi_F-\phi_B)} \right \rangle = 
\frac{1}{N_{\rm events}} \sum_{\rm events} \frac{1}{n_F n_B} \sum_{i=1}^{n_F} \sum_{j=1}^{n_B}  e^{i k(\phi_i-\phi_j)}, 
\end{eqnarray}
with $k$ denoting the Fourier rank and $\phi_i$ ($\phi_j$) being the azimuthal angles of particles emitted in the 
forward (backward) rapidity windows. The quantities $n_F$ and $n_B$ are the
corresponding multiplicities and $N_{\rm events}$ is the number of events.
When no correlations between particles are present, the distribution function of $n$ particles
is the product of one-body distributions
\begin{eqnarray}
f(\phi) = v_0 + 2 \sum_{k=1} v_k \cos[k (\phi-\Psi^{(k)})],
\end{eqnarray}
and one obtains
\begin{eqnarray}
\left \langle e^{i k(\phi_F-\phi_B)} \right \rangle = \left \langle v_{k,F} v_{k,B}  \cos (k \Delta_{FB}) \right \rangle_{\rm events},
\label{eq:cum2}
\end{eqnarray}
where $\Delta_{FB}=\Psi^{(k)}_F-\Psi^{(k)}_B$ is the relative torque angle between the principal axes in the forward and backward directions.
Non-flow (nf) contributions (resonance decays, jets, conservation laws, Bose-Einstein  
correlations, short-range correlations, etc.) modify the right-hand side of Eq.~(\ref{eq:cum2}) at the level $1/n$, where $n$ denotes the 
effective multiplicity. Since we are interested in $\cos [k(\Psi_F-\Psi_B)]$, we divide  
Eq.~(\ref{eq:cum2}) by $v_{k,F} v_{k,B}$ by evaluating the following ratio of cumulants:
\begin{eqnarray}
\cos(k\Delta_{FB})\left \{ 2 \right \} \equiv \frac{\left \langle e^{i k(\phi_F-\phi_B)} \right \rangle }
{\sqrt{ \left \langle e^{i k(\phi_{F,1}-\phi_{F,2})} \right \rangle \left \langle e^{i k(\phi_{B,1}-\phi_{B,2})} \right \rangle} } = 
\left \langle \cos (k \Delta_{FB}) \right \rangle_{\rm events}  +{\rm nf}. \label{eq:c2}
\end{eqnarray}
One may also use higher-order cumulants to generate statistical measures of the torque. For example, 
the ratio of four-particle cumulants yields
\begin{eqnarray}
\cos(2k \Delta_{FB})\left \{ 4 \right \} \equiv
\frac{\langle  e^{i k [(\phi_{F,1}+\phi_{F,2})-(\phi_{B,1}+\phi_{B,2})]} \rangle}
  {\langle  e^{i k [(\phi_{F,1}-\phi_{F,2})-(\phi_{B,1}-\phi_{B,2})]} \rangle}= \left \langle \cos (2 k \Delta_{FB}) \right \rangle_{\rm events}  +{\rm nf}.
\label{eq:c4}
\end{eqnarray}
The important issue in such of studies is the influence of the non-flow contributions.

\begin{figure}[tb]
    \parbox{.5\textwidth}{%   %\def\halftext{.471\textwidth}
    \includegraphics[width=.5\textwidth]{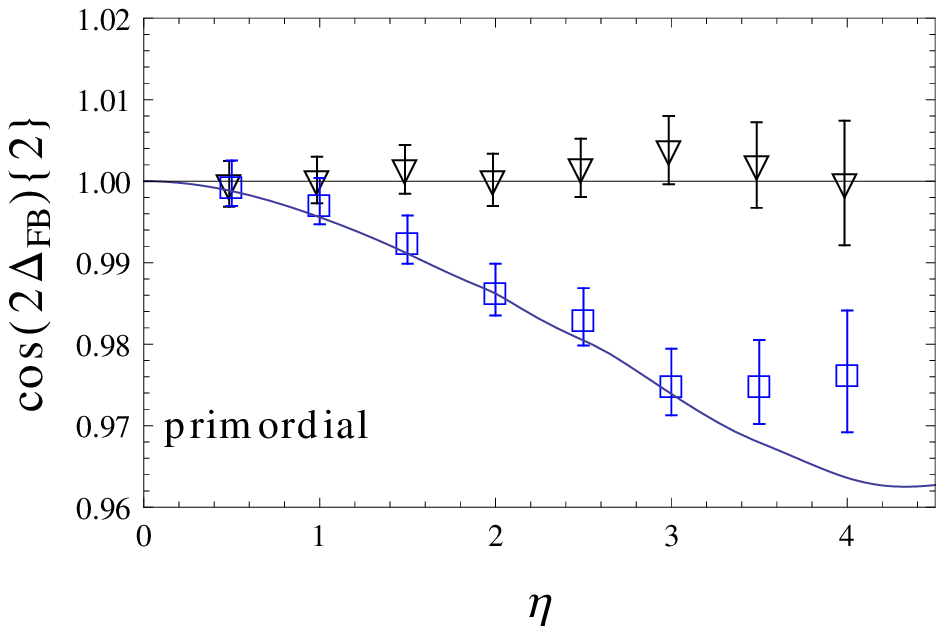}
\caption{{\tt THERMINATOR} simulations (100000 events) for Au+Au collisions at the highest RHIC energy 
for $c=20-25\%$. 
The cumulants are evaluated with the primordial particles only (i.e. with no resonance decays). 
Triangles correspond to the case with no torque, squares to case with the torque. 
 \label{fig:prim}}}
            \hfill
     \parbox{.5\textwidth}{\vspace{3mm}
     \centerline{\includegraphics[width=.5\textwidth]{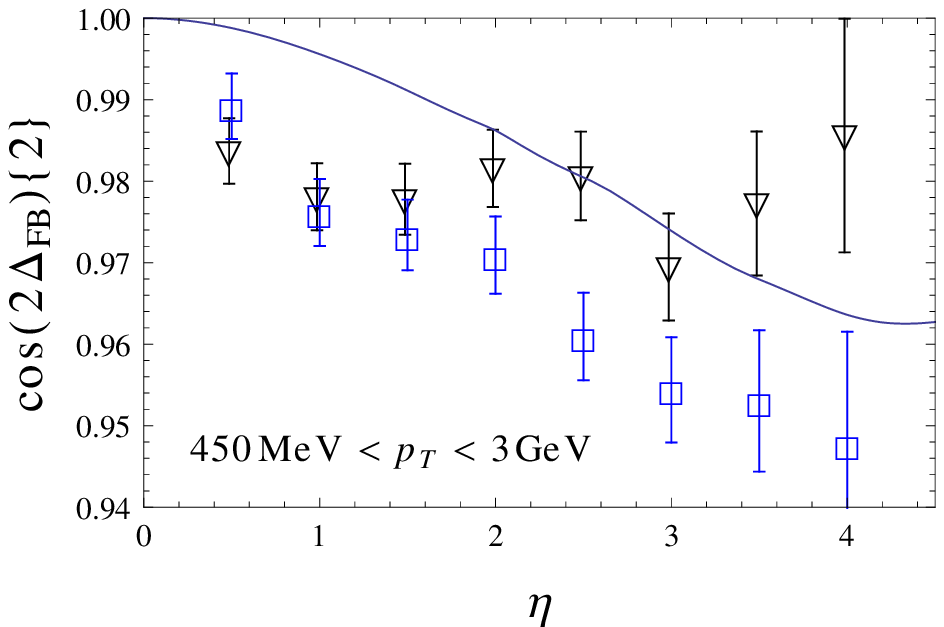}}
\caption{Same as Fig.~\ref{fig:prim} from all final charged pions, kaons, protons, and antiprotons,
with $450~{\rm MeV} < p_T < 3~{\rm GeV}$. The departure of the triangles (no torque) from unity 
displays the non-flow contribution from resonance decays. The torqued case (squares) is visibly shifted from the 
case without the torque (triangles).
\label{fig:all}}}
\end{figure}

We have run {\tt THERMINATOR} \cite{Kisiel:2005hn} simulations on top of the hydrodynamic solutions, taking the case without the torque and with 
the torque, fixing the torque angle to a typical value of 8$^o$ at centrality 20-25\%. The results are displayed in Figs.~\ref{fig:prim} and
\ref{fig:all}.
The solid line represents the fireball torque angle of the velocity distribution after the hydrodynamic evolution. 
The agreement of the line and the squares shows that the statistics is sufficient to detect the torque effect.
We note a sizable departure of the no-torque and torque cases in Fig.~\ref{fig:all}, showing that the effect may be observed 
in experimental data.

In conclusion, we summarize our results: 1)~The space-time rapidity emission profile, where the initial 
longitudinally-moving sources emit predominantly in the 
direction of their motion, combined with the statistical fluctuations of the source densities in the transverse plane, 
lead to event-by event torqued fireballs. 2)~The standard deviation of the relative torque angle in the fireball 
between the forward ($\eta_\parallel \sim 3$) 
and backward ($\eta_\parallel \sim -3$) regions varies from $\sim 20^\circ$ for the most central collisions to $\sim 10^\circ$ 
for the mid-central and  mid-peripheral Au+Au collisions at the highest RHIC energies.
3)~The torque of the initial fireball yields, via hydro evolution, the torque 
of the transverse fluid velocity, and, finally, turns into the torque of the 
principal axes of the transverse-momentum distributions of the produced hadrons. 
4)~We have proposed measures based on cumulants containing particles in different pseudorapidity bins 
can be used to detect the torque effect experimentally. 
Based on {\tt THERMINATOR} simulations we expect that the torque fluctuations should be 
observed in the high-statistics RHIC data.
5)~The statistical noise at hadronization decreases as the product of the square root of the particle 
multiplicity and the flow coefficient,\cite{Bozek:2010vz} hence it is best to look for the 
torque effect in the mid-central classes, such as $c=20-30\%$, 
and with the exclusion of the soft-momentum hadrons to avoid correlations from resonances.
6)~The torque should have a similar size for the elliptic flow and the triangular flow.
7)~Other models of the initial phase (multi-source models) 
should also be investigated in that regard, as emergence of the effect is generic for asymmetric rapidity 
emission profiles.

%\section*{Acknowledgements}

\bigskip

One of us (WB) acknowledges useful conversations with Paul Sorensen, concerning spotting the torque effect in the 
experimental data. 
Research supported by Polish Ministry of Science and Higher Education, grants N~N202~263438 and N~N202~249235, and by 
the Portuguese Funda\c{c}\~{a}o para a Ci\^{e}ncia e Tecnologia, FEDER, OE, grant SFRH/BPD/63070/2009, CERN/FP/116334/201.


\begin{thebibliography}{99}
%%%%%%%%%%%%%%%%%%%%%%%%%%%%%%%%%%%%%%%%%%%%%%%%%%%%%%%%%%%%%
% Some macros are available for the bibliography:
%  o for general use
%    \JL : general journals                 \andvol : Vol (Year) Page
%  o for individual journal 
%    \AJ   : Astrophys. J.           \NC         : Nuovo Cim.
%    \ANN  : Ann. of Phys.           \NPA, \NPB  : Nucl. Phys. [A,B]
%    \CMP  : Commun. Math. Phys.     \PLA, \PLB  : Phys. Lett. [A,B]
%    \IJMP : Int. J. Mod. Phys.      \PRA - \PRE : Phys. Rev. [A-E]     
%    \JHEP : J. High Energy Phys.    \PRL        : Phys. Rev. Lett.
%    \JMP  : J. Math. Phys.          \PRP        : Phys. Rep.
%    \JP   : J. of Phys.             \PTP        : Prog. Theor. Phys.     
%    \JPSJ : J. Phys. Soc. Jpn.      \PTPS       : Prog. Theor. Phys. Suppl.
% Usage:
%  \PRD{45,1990,345}          ==> Phys.~Rev.\ D \textbf{45} (1990), 345
%  \JL{Nature,418,2002,123}   ==> Nature \textbf{418} (2002), 123
%  \andvol{123,1995,1020}    ==> \textbf{123} (1995), 1020
%%%%%%%%%%%%%%%%%%%%%%%%%%%%%%%%%%%%%%%%%%%%%%%%%%%%%%%%%%%%%
  
\bibitem{Bozek:2010vz}
P.~Bo\.zek, W. Broniowski, J. Moreira, \PRC{83,2011,034911}

\bibitem{Bialas:1976ed}
A. Bia\l{}as, M. B\l{}eszy\'nski, W. Czy\.z, \NPB{11,1976,461} 
 
\bibitem{Bialas:2004su}
A. Bia\l{}as, W. Czy\.z, Acta Phys. Polon. B {\bf 36} (2005) 905      

\bibitem{Bzdak:2009dr}
A. Bzdak, K. Wo\'zniak, \PRC{81,2010,034908}

\bibitem{Bozek:2010bi}
P. Bo\.zek, I. Wyskiel, \PRC{81,2010,054902}

\bibitem{Kisiel:2005hn}
A. Kisiel, T. Ta\l{}u\'c, W. Broniowski, W. Florkowski, Comput. Phys. Commun. {\bf 174} (2006) 669;
M.~Chojnacki, A.~Kisiel, W.~Florkowski, W.~Broniowski, arXiv:1102.0273 [nucl-th]
 
\end{thebibliography}
\end{document}